\documentclass[manuscript]{acmart}
\AtBeginDocument{%
  \providecommand\BibTeX{{%
    \normalfont B\kern-0.5em{\scshape i\kern-0.25em b}\kern-0.8em\TeX}}}

\setcopyright{acmcopyright}
\copyrightyear{2026}
\acmYear{2026}
\acmDOI{XXXXXXX.XXXXXXX}

\acmJournal{TIIS}
\acmVolume{1}
\acmNumber{1}
\acmArticle{1}
\acmMonth{1}

\usepackage{times}
\usepackage{helvet}
\usepackage{courier}
\usepackage{graphicx}
\usepackage{color}
\usepackage{caption}
\usepackage{subcaption}
\usepackage{enumitem}
\usepackage{nicefrac}
\usepackage{ulem}
\usepackage{soul}
\usepackage[justification=centering]{caption}

\begin{document}

\definecolor{purple}{rgb}{0.5, 0.0, 0.5}
\newcommand{\note}[1]{\textcolor{red}{#1}}
\newcommand{\TODO}[1]{\textcolor{red}{#1}}
\newcommand{\eric}[1]{\textcolor{blue}{(#1)}}
\newcommand{\nah}[1]{\textcolor{green}{(#1)}}
\definecolor{blue-violet}{rgb}{0.54, 0.17, 0.89}
\newcommand{\donald}[1]{\textcolor{red}{(#1)}}
\newcommand{\removed}[1]{\textcolor{red}{\sout{#1}}}
\newcommand{\added}[1]{\textcolor{black}{#1}}
\newcommand{\minor}[1]{\textcolor{black}{#1}}

\title{Human-in-the-Loop User Feedback Affects Perceived Accuracy and Trust, but Task Subjectivity Matters}

\author{Donald R. Honeycutt}
\email{dhoneycutt@ufl.edu}
\affiliation{%
  \institution{University of Florida}
  \city{Gainesville}
  \state{Florida}
  \country{USA}
}
\author{Mahsan Nourani}
\email{m.nourani@northeastern.edu}
\affiliation{%
  \institution{Northeastern University}
  \city{Boston}
  \state{Massachusetts}
  \country{USA}
}
\author{Eric D. Ragan}
\email{eragan@ufl.edu}
\affiliation{%
  \institution{University of Florida}
  \city{Gainesville}
  \state{Florida}
  \country{USA}
}

% \renewcommand{\shortauthors}{Honeycutt et al.}

\begin{abstract}
\label{sec:abstract}
\added{
While machine learning can produce complex models beyond those that a human could produce manually, incorporating human input can often improve intelligent system performance beyond purely data-driven models.
}
These human-in-the-loop approaches can also be used to dynamically update a model in the presence of shifting data or system goals. 
While this feedback could come from system designers or domain experts, in many cases, the end users who regularly use the system will naturally develop an understanding of its flaws and desire the ability to change the system's behavior based on their knowledge.
While soliciting feedback from end users can result in significant model improvement over time, introducing these feedback techniques can also affect several human factors---such as trust or perception of system accuracy---that are not yet fully understood and have different effects reported in the existing literature.
Therefore, we sought to build on the existing research to further explore how the act of providing feedback can affect user understanding of an intelligent system and its accuracy in different contexts.
We present three controlled experiments that study the effects of interactive feedback collections on user impressions in domains with objective and subjective feedback.
The results show that in a context where there is an objectively correct answer, providing human-in-the-loop feedback lowered both participants' trust in the system and their perception of system accuracy, regardless of whether the system accuracy improved in response to their feedback.
However, when the feedback being provided involved subjective opinion, no such negative bias was observed.
Furthermore, in the objective context, participants distrusted the system over time, whereas participants in the subjective context mistrusted the system over time.
These results highlight the importance of considering the effects of allowing different types of end-user feedback on user trust when designing intelligent systems.
% \donald{Change the results statement to include both experiments and say something about the results showing that context matters for the observed effects or something}
% The results show that providing human-in-the-loop feedback lowered both participants' trust in the system and their perception of system accuracy, regardless of whether the system accuracy improved in response to their feedback.
% These results highlight the importance of considering the effects of allowing end-user feedback on user trust when designing intelligent systems.

\end{abstract}




\maketitle

\section{Introduction}
\label{sec:intro}
%Getting human input in ML models is good
There are many reasons designers of intelligent systems may want to incorporate human input into the training of their \added{machine learning (ML)} models.
Domain experts may have pre-existing understanding of complex application areas that may be usable to address the unreducible error present when learning from data alone.
This process may be more efficient than purely data-driven ML, particularly when there does not already exist sufficient labeled training data and obtaining such labels is prohibitively expensive.
Additionally, human modification can be used to tune model behavior to correct errors and patterns learned from the training data that do not generalize to patterns in the data as a whole.
% Erroneous model behavior caused by changes in data over time can also be corrected by system end users who notice and correct the model's failures.
% \note{citations for stuff above?}
ML systems can support this by utilizing \textit{human-in-the-loop} (HITL) approaches that modify the model to be closer to the feedback provided by a human.
The most basic approach that is prevalent in the existing body of work---new annotation of unlabeled data---uses traditional ML on an updated set of training data to make predictions that are closer to the user's labels, often in the form of corrective feedback where the user identifies and corrects instances where the model makes mistakes or labels data that is identified as statistically likely to result in the greatest model improvement
% ~\removed{\cite{cohn1996active,settles2011theories,hanneke2014theory,baum1992query,smailovic2014stream,cohn1994improving,lewis1994sequential,tong2001support,hwang1991query,seung1992query,fu2013survey,kumar2020active}}
\added{\cite{lewis1994sequential,kumar2020active,cohn1994improving}}.

Feedback is of course not limited to mere annotation, which may be an inefficient method of achieving specific behavioral changes within a model.
Direct modification of model behavior without reliance on retraining with new labeled data allows for a greater level of control in which the user can mold the models behavior to their liking.
While directly modifying feature weights is a simple approach that can be effective in some scenarios, this requires either a simple and inherently interpretable model or a strong understanding of the ML model in use and how to modify it.
Explainability techniques that provide information about the model's behavior can not only help the user understand the model they are trying to modify but can also serve as a method of giving a user intuitive control over model behavior without explicit modification of model parameters.
This approach---\textit{explanatory interactive learning}---has the user provide modifications to the explanations to show how they think the model should behave differently~\cite{teso2019explanatory,kulesza2010explanatory,holzinger2016towards,li2020explanations,ghai2020explainable}; by producing explanations that more closely adhere to the provided modifications, the updated model should more closely represent the mental model of the user.

% \eric{change beginning of intro so not exactly the same as other paper}
% Bringing human feedback into the development of machine learning models has many benefits.
% At its simplest, human feedback allows a model to incorporate new annotations for unlabeled data to increase performance by improving the training set.
% A common method for introducing human feedback is active learning, where the selection of data to obtain labels for is left to the model~\cite{cohn1996active}.
% Alternatively, a more human-centered approach has the labeler choose which instances to be labeled, relying on human intuition to decide what feedback would be most relevant to improve the model based on observations of its performance~\cite{tong2001support}.
% Developers can also allow for further involvement by giving the human participant feature-level control over model parameters, such as allowing direct modification of the feature space and its associated weights~\cite{cho2019explanatory} or prioritizing decision rules used by the model~\cite{yang2019study}.

%End users of systems can be involved in the feedback loop
Frequently, the HITL approaches either involve system developers for development and debugging~\cite{vathoopan2016human} or independent workers on crowd-sourcing platforms~\cite{li2017human}.
By taking advantage of end-users' periodical feedback upon noticing errors, these models can stay updated in the presence of shifting data or changing goals~\cite{geng2009incremental,yamauchi2009optimal,elwell2011incremental}.
Systems can also update over time by implicitly capturing user behaviors, which is a technique commonly used in recommender systems~\cite{shivaswamy2012online,middleton2003capturing}.
While this feedback is not provided explicitly, users can still observe the system directly reacting in response to their actions, decisions, and feedback.
Furthermore, end users of intelligent systems may want the ability to correct observed model errors.
When engaged with the outcomes of a system, many users desire the ability to influence those outcomes by providing feedback beyond simple error correction~\cite{stumpf2008integrating}.

%T
While HITL systems can have improved model accuracy and provide users control over the systems they rely on, there may also be unexplored consequences to allowing end users to provide feedback.
For instance, Van den Bos et al.~\shortcite{van1996consistency} observed that when interacting with human teams, the ability to provide feedback has been observed to have a positive effect on the perceived fairness of team decisions.
In their study, users who felt their feedback was considered reported higher levels of trust in the decision-making process and were more committed that the correct decision was made.
They also observed the inverse effect, with a decrease in trust in the team if feedback was provided but ignored~\cite{korsgaard1995building}.
Since providing feedback to an automated decision-making system is similar to providing feedback to a human-based decision making system, it is possible that similar effects could be observed in HITL systems.

\added{
If providing feedback does affect user trust, it could lead to people misusing the systems they provide feedback to.
When experiencing a higher level of trust than is appropriate based on the system performance, users may over rely on the system.
On the other hand, having a lower level of trust could result in not using the system at all~\cite{lee2004trust,parasuraman1997humans}.
Therefore, it is important to understand how providing feedback to an intelligent system affects trust so that it can be accounted for when designing HITL systems.
Prior research has shown that the presence of HITL elements in intelligent systems can have both positive and negative effects on user perception and trust.
Users are consistently shown to have both stated preferences towards systems with HITL elements~\cite{parra2015user,jin2017different,stumpf2008integrating} and are more willing to use and rely on HITL systems than non-HITL intelligent systems~\cite{dietvorst2018overcoming,parra2015user}.
However, such mechanisms can result in increased cognitive load that causes decreased engagement with the system~\cite{jin2017different,ehrmann2022evaluating}.
In systems where the HITL feedback is purely correcting objective system errors, it can result in underestimation of accuracy and distrust~\cite{honeycutt2020soliciting,iuzzolino2020automation}.
While prior research has clearly established the diametric nature of HITL feedback, further research is needed to understand the reasons behind the differences in observed effects in different contexts.
Understanding the effect that providing end users with HITL feedback mechanisms will have in a given system and context is important to ensure it does not have an adverse effect on their perception and use of the system.
}

% \removed{In this paper, we present the results of three experiments that build on our prior study~\cite{honeycutt2020soliciting} that showed a negative bias towards perceived system accuracy and trust among those who provided interactive human-in-the-loop feedback compared to those who observed the same system outputs but did not provide feedback.} 
\added{
In this paper, we present the results of three experiments that build on our prior study~\cite{honeycutt2020soliciting} that showed participants who provided HITL feedback underestimated system accuracy and had reduced trust compared to those who did not.
These experiments aim to explore the effects of providing HITL feedback in both subjective and objective contexts.
}
This paper is a direct extension of the previously published conference paper on the previous study. 
In this extension, the previously published experiment from~\cite{honeycutt2020soliciting} was expanded in scope with new experimental conditions to allow a deeper analysis; the expanded study is presented as Experiment 1 of this paper. 
In addition, Experiments 2 and 3 are entirely new.

Experiment 1 seeks to isolate the difference between the level of feedback interactivity and the belief that the system is updating based on user feedback.
We used a simulated object-detection system---the same system as used in prior work~\cite{honeycutt2020soliciting}---that tasked users with providing interactive feedback via adjusting bounding boxes for detected objects in images.
Additionally, this experiment also controlled for whether participants were led to believe that their feedback was resulting in dynamic system changes throughout the task---despite no such updates being made.
We found that belief that the system was updating had no significant effect on perception of system accuracy or user trust.
However, those who provided interactive feedback perceived the system to be significantly less accurate than those who only provided basic feedback about whether the classification was accurate or not.
Similarly, trust in the system was lower among those who provided interactive feedback compared to participants who did not.
Regardless of condition, all participants found the system to become less accurate over time, despite the true accuracy of the system remaining constant.

Our second and third studies recreate the experimental design of the first study, but this time in the context of text classification instead of object detection.
In this context, the interactive feedback comes from adjusting the list of words which are the most relevant towards identifying the text sample's topic.
Where adjusting bounding boxes in images has an objectively correct answer---either there is an object in a location or there is not---the most relevant words to identify the topic of a paragraph is somewhat subjective.
We hypothesized that this difference would re-contextualize the act of providing interactive feedback from having to correct system errors to making the system's model behave more closely to the user's mental model.
Since shaping system behavior to be more similar to one's own mental model is a reason users have provided for preferring HITL systems in prior research~\cite{rani2022investigating,parra2015user}, we hypothesized that this re-contextualization would help alleviate the negative biases observed in the image classification study.
Unlike the image classification context, we did not observe any effects on trust or perception of accuracy based on level of feedback interactivity.
Also contrary to the results in the image classification context, participants in all conditions perceived the system to increase in accuracy over time, even though no such accuracy change occurred. 

\added{
Additionally, in this paper we discuss the existing literature showing both positive and negative effects of HITL feedback on user perception and the differences in system and task context present in each of those studies, particularly as it relates to feedback subjectivity.
This discussion---in combination with the results of the studies presented in this paper---contributes an improved understanding of why different studies of the effects of providing HITL feedback have produced contrasting results.
}

\section{Related Work}
\label{sec:related}
The presented research builds on prior work from the perspectives of HITL ML and trust in artificial intelligence.
% \eric{would expect related work to be expanded for a journal paper. this looks short. i expect you should be able to incorporate some content from your qual or proposal to make this easier}

%Active learning/human-in-the-loop/relevance feedback (usefulness of human feedback in ML models)
\subsection{Human-in-the-Loop Machine Learning}

While ML can be used to train models based purely on data without direct human guidance, there are many scenarios where incorporating human feedback is beneficial.
In many cases, this feedback is simply having a human annotate new data to be incorporated into the model.
While data annotation can be viewed as merely a requirement to obtain the large pool of labeled data necessary to train a traditional supervised learning model, data annotation also serves as an effective medium for people to transfer their knowledge into a complex modeling process.
This approach allows people who may be completely inexperienced with ML to provide feedback to update a model using their understanding of the domain.
As data annotation can be time consuming and expensive, much research has been performed on how to optimize the selection of which data to label.
\textit{Relevance feedback} is a HITL method where a human reviews the pool of unlabeled data alongside the current model's predictions on that data, choosing when to provide new labels to the system based on their own intuition~\cite{tong2001support}.
As human intuition does not result in optimal selections of what data to label in many domains~\cite{cakmak2014eliciting}, another approach is to choose instances to add to the training set using objective metrics based on the model.
\textit{Active learning} selects relevant instances to show to a human---referred to as an \textit{oracle}---based on which unlabeled data are most likely to represent information missing in the current version of the model~\cite{cohn1996active}.
While theoretical active learning research treats the oracle as merely being a way to obtain the true labels for selected data, in practice, active learning models need to account for the fact that the oracle is a human and therefore not infallible~\cite{settles2011theories}.

%Interactive learning (types of human feedback beyond simple annotation)
\added{
While data annotation and error correction are a simple medium for providing HITL feedback to the model, it may be more efficient to give richer feedback that allows for more direct modification of model parameters, allowing the oracle to mold the ML system to align with their own mental model of the task and data set.
The most direct method is to allow for direct control of model parameters, such as features and their weights~\cite{cho2019explanatory} or the rules that are used to make decisions within the model~\cite{yang2019study}.
Being able to control model parameters in this way has been found to be useful for debugging models~\cite{kulesza2010explanatory}.
However, while direct parameter modification can be effective, the complex nature of black-box models often causes those techniques to be ineffective or inefficient, particularly among users who are inexperienced in ML.
When the level of user control is too complex for the user, it can result in an increase in cognitive load and lower engagement with providing feedback than if they had less control~\cite{jin2017different,ehrmann2022evaluating,rani2022investigating}.
A potential approach to allow for rich user control through an understandable medium is to incorporate explainable AI techniques and have feedback be based on those explanations.
}
Explanatory interactive learning has the oracle not only provide the appropriate label for the data point but also provides an explanation of the current model prediction and asks the oracle to correct the reasoning in the explanations~\cite{teso2019explanatory}.
This helps avoid situations where the model has a flaw that happens to result in the correct prediction by chance.
Examples of explanation types that can be used for interactive feedback include highlighting relevant words for text classification~\cite{kulesza2015principles,kulesza2010explanatory,zaidan2007using}, ranking features in tabular data~\cite{cho2019explanatory}, and bounding boxes for image classification and object detection tasks~\cite{fails2003interactive}.
While this higher level of control over the model may not be desirable in all applications, Holzinger et al.~\shortcite{holzinger2016towards} showed that human-machine teaming can sometimes result in a closer to optimal model than ML alone.

Another approach to creating interactive explanations to support human feedback is the use of interpretable surrogate models to generate approximate explanations of model behavior that are applicable to any model architecture~\cite{roy2021explainable,zhao2021baylime,torcianti2021explainable,vcik2021explaining}.
The main challenge in using surrogate explainer models for HITL systems is in updating the original model to be closer to the explanations provided by the oracle when the explanations do not directly correspond to the format of model inputs.
A simple strategy is to use the updated surrogate model to generate a robust set of counterexamples to the data set to retrain the original model with new data that are representative of the changes made by the oracle~\cite{yang2019study,qian2019systemer}.
By adding these counterexamples to the training data, both performance and explanation quality can be improved when compared to regular active learning~\cite{teso2019explanatory}.

%Incremental learning/concept drift/coactive learning (benefits of end user feedback/continuously updating the model after its in use, reason that trust of people providing feedback is relevant)
While the person providing feedback is not necessarily the end user for many HITL systems, there are advantages to bringing end users into the loop.
Stumpf et al.~\shortcite{stumpf2008integrating} found that users of intelligent systems largely want to provide feedback to systems they are using, particularly when it gives them a feeling of being able to control some aspect of the model.
Similarly, people are more likely to use an imperfect intelligent system when they have the ability to correct its errors \cite{dietvorst2018overcoming}.
Additionally, end users may notice when an already deployed system begins to falter.
Even if a model was very accurate at the initial time of training, the training data may become less representative of the actual population of data as trends shift over time.
This is a phenomenon known as concept drift~\cite{vzliobaite2010learning}.
A HITL approach to dealing with this problem is known as \textit{incremental learning}, where the model periodically obtains labels as they become available to update the model while it is in use~\cite{geng2009incremental}.
These techniques have been shown to effectively address the problem of concept drift in ML systems~\cite{yamauchi2009optimal,elwell2011incremental}.

%Transition into trust and how providing feedback could affect how users perceive systems (effects of usage of input on trust in the psychology of collaborative decision making)
Providing input has been shown to affect trust and perception of fairness in the field of psychology.
In decision-making teams, people were observed to place more trust in a team-leader who actively considered their input, and they were also more confident that the correct decision was made after the fact~\cite{korsgaard1995building}.
A similar effect was observed in procedural decision making systems, with people having a higher level of trust and perception of fairness in a decision-making system that they were able to give input to~\cite{van1996consistency}.
An interesting result from both of these studies was that providing feedback had a negative effect on trust if the feedback was ignored.
Since feedback affects interpersonal trust by improving trust if feedback is considered and decreasing trust if it is ignored, a similar effect may be observed in human-computer interactions.
When users of HITL systems cannot directly see how their feedback has changed the model, they can become frustrated and be unsure of what to do next~\cite{stumpf2008integrating,kulesza2009fixing}.
This is not only an issue of a lack of detection of model changes, but also that users may expect large changes to happen faster than is feasible for a given architecture.
Providing information about how a person's contributions are being used has been shown to increase participation in social groups~\cite{rashid2006motivating}, which may motivate similar information being provided to explain how feedback is being used in HITL systems to both explain effects and improve labeling quality~\cite{amershi2014power}.
\added{
Additionally, when users are treated as an active learning oracle and are asked to correct objective system errors, it can result in them distrusting the system~\cite{honeycutt2020soliciting,iuzzolino2020automation}.
However, HITL features that allow direct control over model parameters to better match personal preferences can result in increased satisfaction and willingness to use the system~\cite{rani2022investigating,parra2015user,dietvorst2018overcoming}.
While many papers have shown that the presence of HITL feedback can have both positive and negative effects on user trust and perception of intelligent systems, they use a variety of contexts and the reasons behind the differing results have not been sufficiently established.
}

\subsection{Trust in Artificial Intelligence}
User trust in artificial intelligence systems has been studied for many years and is of value since it is directly associated with usage and reliance~\cite{parasuraman1997humans,siau2018building}.
As a result, users need to place an appropriate amount of trust in a system based on its performance in different contexts.
Reliance and trust in automated systems are not binary (i.e., to trust or not) and are generally more complex~\cite{lewicki1998trust,lee2004trust}.
Desired behavior is for a user to examine a system's outputs and decide whether to rely on the system based on the accuracy of results~\cite{hoffman2013trust}.
This behavior has been observed to be more prevalent among users who are domain experts than novice users~\cite{nourani2020Role}.
Sometimes, however, users might trust a system completely without checking the outcomes, i.e., over-reliance or \textit{automation bias}~\cite{goddard2012automation}.
This situation can be caused by a user's lack of confidence or when the system seems more intelligent than they are  based on their initial preconceptions~\cite{lee2004trust,hoffman2013trust,nourani2020investigating}.
In contrasting scenarios, \textit{mistrust}~\cite{parasuraman1997humans} and \textit{distrust}~\cite{lee2004trust} can cause users to rely more on themselves and avoid using a system they perceive to be more flawed than it is~\cite{dietvorst2015algorithm}.
Both of these situations can be dangerous, especially for systems with critical tasks where decisions can be fatal.
\added{
For example, wrong decisions in criminal forecast systems can wrongfully convict an innocent person~\cite{berk2015machine} and miscalibrated trust in an AI healthcare system can result in misdiagnosis~\cite{wong2025trust}.
The required level of trust and the trust calibration process also differ greatly by task; cooperative tasks such as decision support systems require the user to  calibrate how often they utilize the system on a case-by-case basis, whereas delegative tasks such as autonomous driving systems require the user to choose whether or not to trust the system as a whole~\cite{wischnewski2023measuring}.
}

\added{
To help calibrate users' trust to an appropriate level and provide more information to aid them in their decision-making process, researchers have explored the use of explainability in artificial intelligence systems~\cite{ribeiro2016should,doshi2017towards}.
}
Perceived transparency due to system explanations contributes to an increase in user trust~\cite{stephanidis2019seven,szczuka2024let} and can encourage usage of an artificially intelligent system~\cite{hoff2015trust}.
Studies of HITL paradigms have shown explainability can help users understand and build trust in the algorithms in order to provide proper feedback and annotations~\cite{ghai2020explainable,teso2018should}.
Explanations of system behavior improve users' mental models of the system's inner workings~\cite{kulesza2013too}, which is strongly linked with developing an appropriate level of trust in a system~\cite{hoffman2021measuring}.
However, if the provided explanations are insufficiently meaningful to the user it can result in an underestimation of system accuracy~\cite{nourani2019effects}.
Humans are additionally prone to several cognitive biases when interacting with intelligent systems~\cite{nourani2019effects,nourani2024user}; system explanations can both help mitigate or exacerbate the effects of such cognitive biases~\cite{bertrand2022cognitive}.

Researchers use different methods to measure trust and reliability in ML and artificial intelligence systems.
Trust in automation is a multi-faceted concept, defined by Lee and See~\cite{lee2004trust} as "the attitude that an agent will help achieve an individual's goals in a situation characterized by uncertainty and vulnerability."
To capture this attitude, there has been much research into survey scales that measure different aspects of trust~\cite{madsen2000measuring,cahour2009does,jian2000foundations,hoffman2021measuring}.
For example, Madsen and Gregor~\cite{madsen2000measuring} break down trust in automation into five constructs: perceived reliability, perceived technical competence, perceived understandability, faith, and personal attachment.
In addition to measuring trust subjectively through questionnaires, there are many behavioral metrics that are highly correlated with trust.
For example, some researchers utilize user's agreement with the system outputs as a measure of reliance and trust;
specifically, identifying when the user agrees with the system outputs that are not correct~\cite{nourani2020don}. 
Yu et al.~\shortcite{yu2019trust} propose a \textit{reliance rate} based on the number of times the users agreed with the system answers out of all their decisions.
In recent work, Yin et al.~\shortcite{yin2019understanding} found that trust is directly affected by user's estimations of the system's accuracy, where underestimation of accuracy can cause mistrust in the system, and vice versa.
As a result, a user's estimated or observed accuracy can be used as an indirect measurement for user trust, and we use these methods in the study reported in this paper.

% \section{Study 1}
% \label{sec:study_1}
% \input{3.study_1}

\section{Experiment 1: Object Detection in Images}
\label{sec:image_study}
\begin{figure*}[ht]
    \centering
    \includegraphics[width=0.99\textwidth]{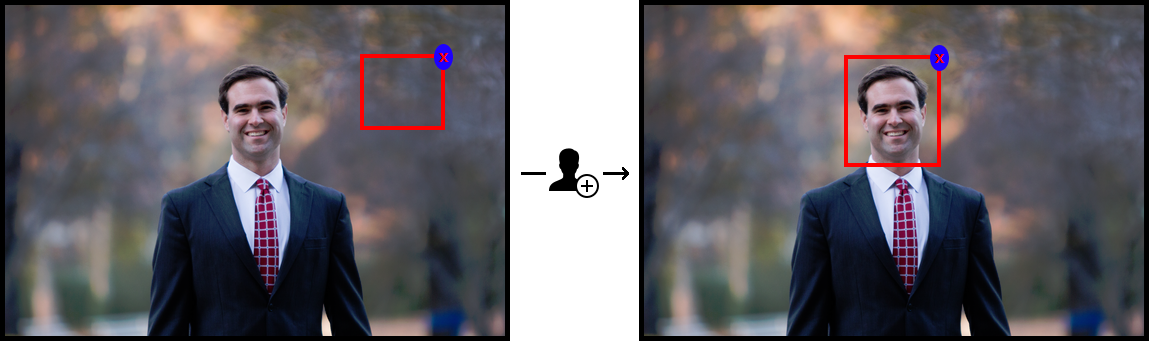}
    \caption{
    In the \textit{interactive feedback} conditions of Experiment 1, participants could delete existing bounding boxes or click-and-drag to create new ones. In this example, the left image shows a system error, and the right shows a version after interactive correction.\protect\footnotemark
    }
    \label{fig:interface}
\end{figure*}

In this section, we discuss the first experiment based around understanding the differences in user trust and perception of system accuracy among users of HITL systems with interactive feedback mechanisms.
We note that part of this experiment was previously published in~\cite{honeycutt2020soliciting}.
We have since expanded the study design by adding additional experimental conditions that will be explained further later in this section. 
The expanded study is reported here with new analyses and results.

\subsubsection{Research Objectives}
With the goal of understanding the relationship between belief that the system is dynamically updating based on user feedback and the level of feedback interactivity, we present the following research questions:
\begin{enumerate}[align=left, label=\textbf{RQ\arabic*:}, ref=RQ\arabic*]
    \item \label{RQ1} Does belief that user feedback is being used to dynamically update system behavior affect user trust or perceived system accuracy?
    \item \label{RQ2.1} Does providing more interactive feedback result in a difference in user trust or perceived system accuracy when compared to only providing accuracy-based feedback?
\end{enumerate}

To address these research questions, we designed a controlled experiment using a simulated object detection system with varying levels of feedback interactivity and a representation of feedback usage.
Based on the results of prior research~\cite{honeycutt2020soliciting}, we hypothesized that the more interactive feedback method would result in a more negative perception of system accuracy and lower levels of trust due to an increased saliency of system errors by spending more time correcting them.
We also hypothesized that no effect would be observed based on the participants being told the system was updating based on their feedback as the provided corrections were of an objective nature rather than being about making the model behave closer to the user's own mental model.
\added{The original study design from~\cite{honeycutt2020soliciting} had change in accuracy as an independent variable, where the system accuracy either improved, stayed the same, or degraded across the three rounds of the task.
As the change in accuracy condition did not result in any significant findings, we chose to eliminate this variable to have a more focused study design.
All study conditions used a constant 80\% system accuracy across all three rounds of the task.}
\footnotetext{\minor{Image from ``Josh McMahon Portraits - 2517'' by John Trainor (used under CC BY 2.0) with annotations added by the authors. License available at https://creativecommons.org/licenses/by/2.0/}}
% % Our initial study design as described above did not fully isolate the effects of providing interactive feedback and the belief that the system would update based on feedback provided.
% Participants either provided binary "yes/no" feedback with the premise of the data being sent to researchers, or they gave interactive feedback---adjusting bounding boxes---under the impression that the system was updating based on that feedback during the task.
% To isolate how each of these components may have contributed to the effects observed in our first study, we created a follow-up study to augment the collected data to address this issue.
% As the change in accuracy condition did not result in any significant findings, we chose to only use a constant accuracy level across the three rounds of the task.

\subsection{Method}
In this sub-section, we present details of our experimental design and study procedure.
\subsubsection{System Design}
\added{We note that the system design for Experiment 1 had no changes to the system design from \cite{honeycutt2020soliciting}.}
We provided participants with a series of images with classifications from a simulated model with detection of human faces as the classification goal.
To allow for interactive feedback, the simulated system outputs include bounding boxes over the location of each detected human face.
\added{
For conditions that involved interactive feedback, bounding boxes could be deleted by clicking an X attached to each bounding box, and new bounding boxes could be drawn by clicking and dragging on the image in the interface as shown in Figure \ref{fig:interface}.
}
While participants were told the classifications came from an artificially intelligent system, the classifications were actually hand-crafted for experimental control purposes.
\added{
Prior research showing user preferences for systems with HITL user control have the limitation that the version of the system that allows for user control could be preferred simply because feedback resulted in the system performing better, not necessarily because of the presence of HITL feedback mechanisms~\cite{rani2022investigating,parra2015user,jin2017different}.
To avoid this potential confound, we chose to use a system that deceives participants to believe that it is updating based on their feedback when in reality all participants viewed the same system outputs regardless of their feedback.
}
% The task consisted of reviewing three rounds of images, with 30 images in each round.

\added{
Since this study uses estimation of system accuracy as one of the main metrics, we needed task rounds that were sufficiently long to ensure participants saw enough system outputs to both make an informed estimation and not just remember the exact percentage of system outputs that were correct.
We additionally needed enough rounds of the task to be able to detect change over time.
However, these needs had to be balanced by the time constraints of an online user study where participants may lose engagement and provide worse feedback over time.
In consideration of balancing these tradeoffs, we selected a task consisting of reviewing three rounds of images, with 30 images in each round.
}
The images used in our simulated model were taken from the Open Images dataset~\cite{OpenImages,OpenImages2} with our own manually generated annotations.
Each round of 30 images contained 20 pictures of people, with the remaining 10 images containing things such as animals or empty scenery.
For images where we chose to simulate system errors, we used a roughly equivalent mix of false positives (bounding boxes placed on objects that were not human faces) and false negatives (unidentified human faces).

\begin{figure*}[ht]
    \centering
    \includegraphics[width=0.99\textwidth]{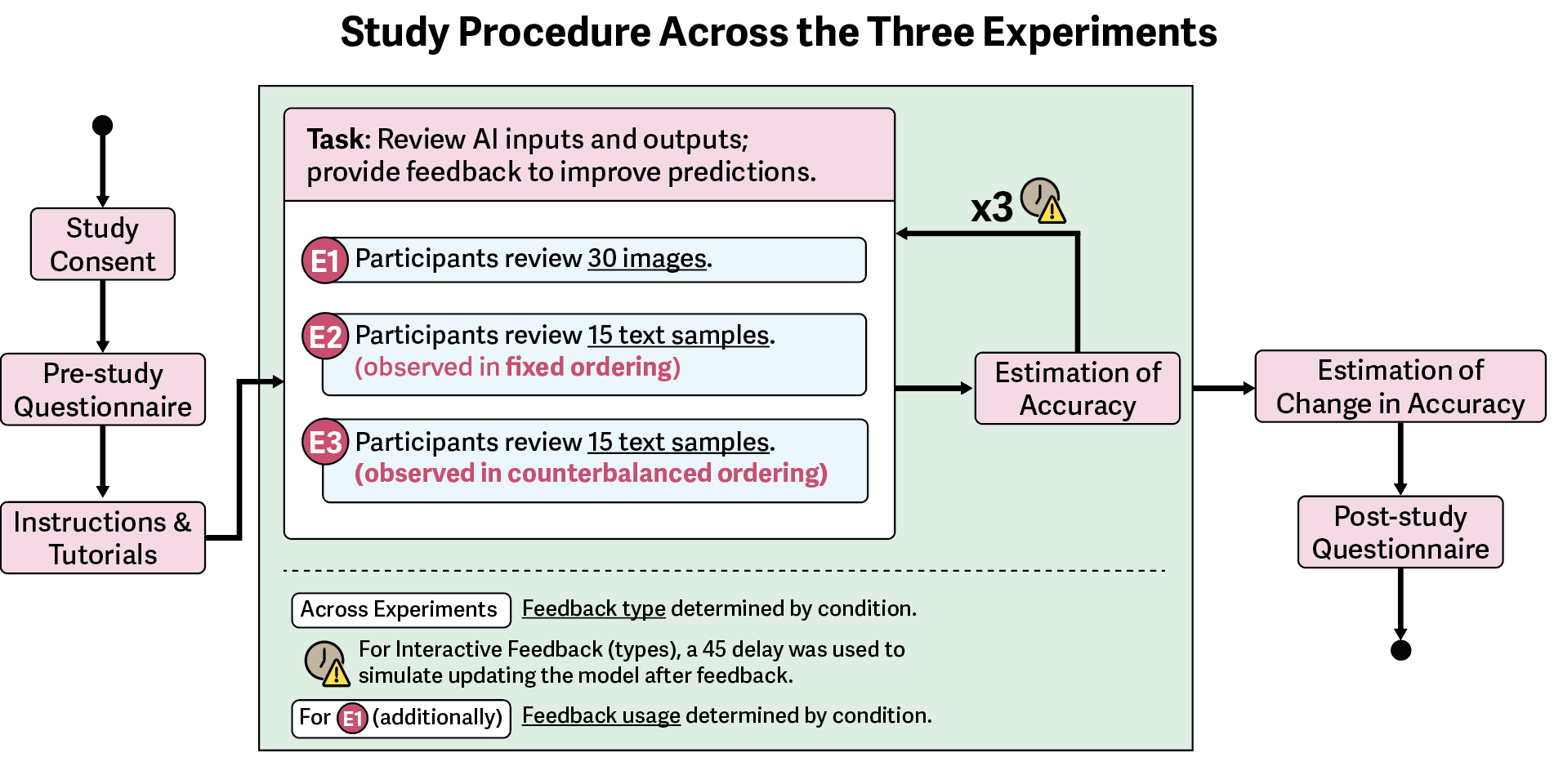}
    \caption{
    \minor{
    An overview of the experimental procedure shared across the three experiments. Each participant interacts with only one experiment. After each round of the task, participants rate system accuracy to capture how perception of accuracy changes over time. This is repeated for three task rounds before moving on to post-study questionnaires.
    }
    }
    \label{fig:flowchart}
\end{figure*}

\subsubsection{Experimental Design}
%  Longer description of system
Because our main metrics---perception of model accuracy and user trust---are based on participants' experiences with the system, we decided that each participant should only see one version of the system so as not to be biased by their experience with the previous system versions.
For this reason, we used a 2x2 between-subjects design for the experiment.
The first independent variable---\textit{feedback type}---pertains to the feedback mechanism used by the participants during the task.
\added{
Those with \textit{binary feedback} responded simply whether the image was classified correctly or not, while those with \textit{interactive feedback} additionally corrected any errors by adding and removing the bounding boxes on top of the images as shown in Figure \ref{fig:interface}.
In this context, the feedback being provided is objective---there is either a human face present in a location or there is not; there is no subjective interpretation of the correct classification.
}
Our second independent variable---\textit{feedback usage}---was based on the information provided to the participants about what was done with the feedback they provided.
\added{
This variable did not change the nature of how participants interacted with the system at all, but rather changed only the explanation of how the system would be using their responses.
For participants in the \textit{without update} condition, they were told that their responses would be saved and sent to researchers after the task for quality control purposes.
The \textit{with update} participants were explicitly informed that the system would update in-between each of the three task rounds based on the feedback they provided before making the next set of classifications.
}
As before, the system accuracy was actually constant and did not update based on feedback for the purposes of experimental control.
To remind the participants and add to believability, a 45 second pause was present between each round with a message telling participants that the system needed time to update its parameters before they could move on.
\added{
As a method of confirming that participants in \textit{with update} conditions actually believed their feedback was being used, we additionally asked them to rate their agreement with the statement "The system correctly uses my feedback" on a 7-point Likert scale at the end of the task. 
The mean response to this question was 4.71 (slightly positive) with a standard deviation of 1.56, confirming that on average participants in \textit{with update} conditions believed the system was actually updating based on their feedback.
}

\added{For the conditions that mirrored conditions from \cite{honeycutt2020soliciting}---\textit{binary feedback without update} and \textit{interactive feedback with update}---the original data was retained and no new data was collected.
New participant data was collected for the new conditions---\textit{binary feedback with update} and \textit{interactive feedback without update}---and analysis was performed on the combined dataset from the original study and the newly collected data.}

\subsubsection{Participants}

Participants were recruited from Amazon Mechanical Turk with a requirement for participants to have the Masters qualification, an approval rate of greater than $90\%$, and $500$ or more prior tasks completed successfully.
Participants ranged from ages 26--66 and lived in the United States at the time of study completion.
To ensure the quality of participant responses, we measured the percentage of responses for which participants correctly identified whether an image corresponded to a system error or not.
As a quality check, participants were not included in the results if they had less than $75\%$ accuracy for either correct instances or system errors.
Our study had a total of 111 participants, and 4 were removed based on the accuracy criteria.
The remaining 107 participants consisted of 45 females and 62 males.
Participants took approximately 11 minutes on average to complete the study.

\subsection{Experiment 1 Results}
\label{sec:results15}
In this section, we present the measures of the follow-up study and its empirical results.
We report statistical test results along with generalized eta squared ($\eta^2_G$) for effect sizes of ANOVA tests and Cohen's d ($d_s$) for effect sizes of post-hoc tests.

\begin{figure*}[ht]
    \centering
    \includegraphics[width=0.99\textwidth]{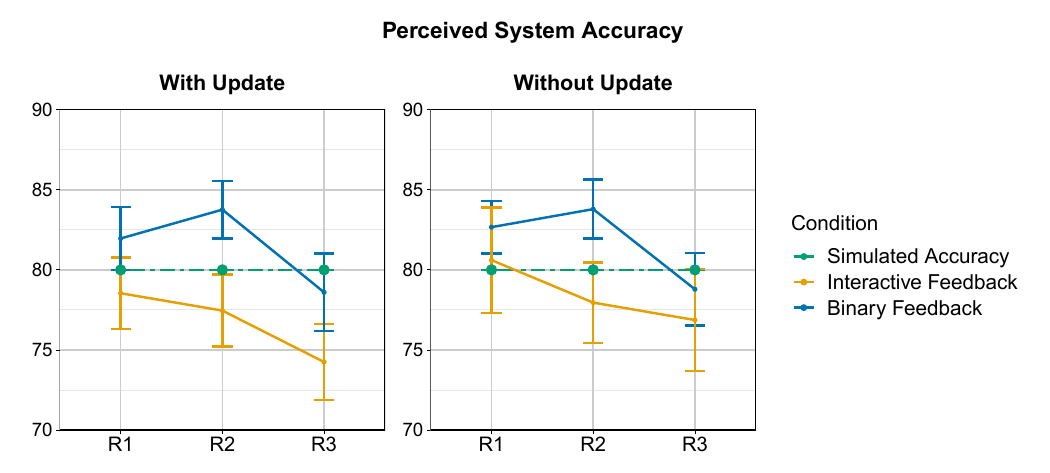}
    \caption{Perceived system accuracy (Percentage) across the three task rounds (R1, R2, R3)}
    % \footnotemark 
    \label{fig:ELine}
\end{figure*}

\subsubsection{User-Perceived Model Accuracy}
Between each round of the task, participants estimated how accurate they thought the system was as a percentage.
\added{They were instructed to make their best estimate of how accurate the system was based on the prior round of the task without directly counting the correct and incorrect instances.
Estimation of system accuracy has been shown to directly affect trust in ML models~\cite{yin2019understanding} and is an established method for estimating general user understanding of an intelligent system's performance and capabilities~\cite{mohseni2018survey,nourani2020don,nourani2020investigating}.}
We performed a three-way mixed-design ANOVA on the estimated accuracy, with \textit{feedback type} and \textit{feedback usage} as between subjects factors and task round (i.e., first, second, or third round) as a within subjects factor.
The analysis showed no significant effects based on \textit{feedback type} ($F(1,103)=3.71, p=0.050$) or \textit{feedback usage} ($F(1,103)=0.17, p=0.676$).
\textit{Task round} was found to have a significant effect on estimated model accuracy with $F(2,206)=11.88$, $p<0.001$, $\eta_G^2 = 0.022$.
Post hoc pairwise t-tests with Bonferroni correction revealed a significant difference between the first and third round with $p<0.050$, with the perceived accuracy in the third round being lower than the perceived accuracy in the first round.
\added{No significant interaction effects were observed between any variables.}
Results are shown in Figure \ref{fig:ELine}.

% \begin{figure}[tb]
%     \centering
%     \includegraphics[width=0.60\columnwidth]{figures/ELine.pdf}
%     \caption{Experiment 1: Perceived system accuracy (Percentage) across the three task rounds (R1, R2, R3). \eric{need to play with design. make easier to interpret}}
%     \label{fig:ELine}
% \end{figure}

% \begin{figure}[tb]
%     \centering
%     % The whole row is forced onto one line
%     \makebox[\textwidth][c]{
%         % Left plot
%         \begin{subfigure}{0.38\textwidth}
%             \centering
%             \includegraphics[width=\linewidth]{figures/ELineUC.pdf}
%             \label{fig:linePredTurk}
%         \end{subfigure}%
%         \hfill
%         % Right plot
%         \begin{subfigure}{0.38\textwidth}
%             \centering
%             \includegraphics[width=\linewidth]{figures/ELineNC.pdf}
%             \label{fig:lineRelTurk}
%         \end{subfigure}%
%         \hfill
%         % Legend
%         \begin{subfigure}{0.20\textwidth}
%             \centering
%             \includegraphics[width=\linewidth]{figures/ELineLegend.pdf}
%             \label{fig:legend}
%         \end{subfigure}%
%     }
%     % Optional global caption
%     \caption{Experiment 1: Perceived system accuracy (Percentage) across the three task rounds (R1, R2, R3.\donald{plot title doesn't have "Perceived System Accuracy", couldn't figure out an easy way that didn't suck}}
%     \label{fig:acc}
% \end{figure}

\subsubsection{Perception of Model Change}
\begin{figure}[tb]
    \centering
    % The whole row is forced onto one line
    \makebox[\textwidth][c]{%
        % Left plot
        \begin{subfigure}{0.39\textwidth}
            \centering
            \includegraphics[width=\linewidth]{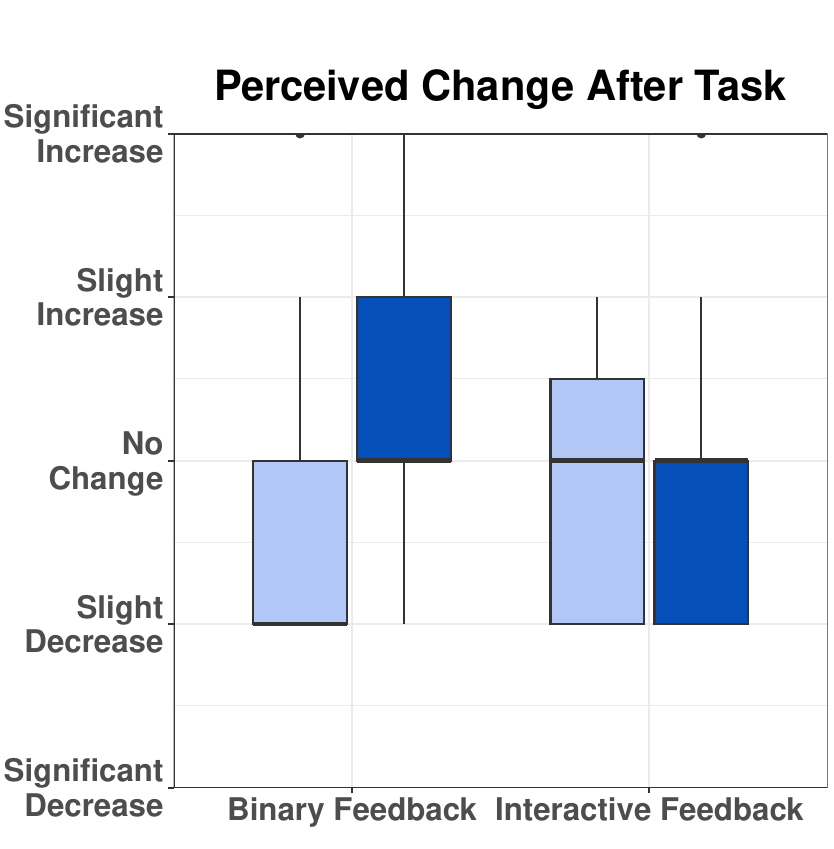}
            \caption{Responses to the question "Did you notice any changes to system accuracy across the rounds of the task?" on a 5-point Likert scale.}
            \label{fig:changeScaleE}
        \end{subfigure}%
        \hfill
        % Right plot
        \begin{subfigure}{0.38\textwidth}
            \centering
            \includegraphics[width=\linewidth]{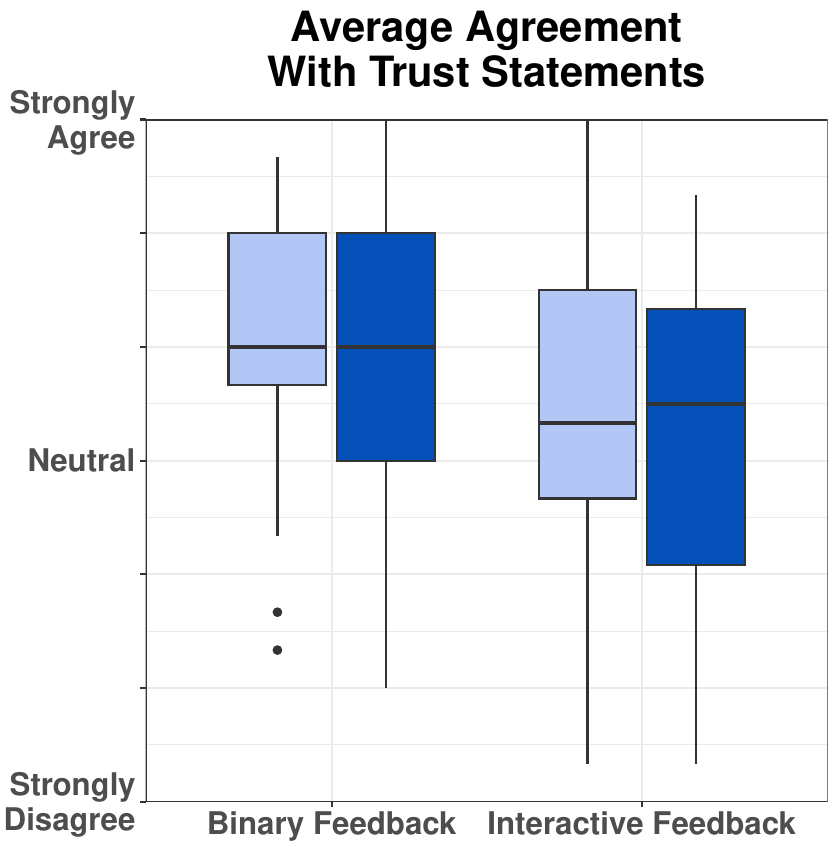}
            \caption{Averaged responses to the three trust scale statements described in Section \ref{subsec:trust1} on a 7-point Likert scale.}
            \label{fig:trustAvgE}
        \end{subfigure}%
        \hfill
        % Legend
        \begin{subfigure}{0.17\textwidth}
            \centering
            \includegraphics[width=\linewidth]{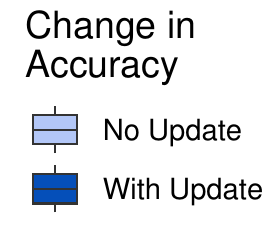}
            \label{fig:legend}
        \end{subfigure}%
    }
    % Optional global caption
    \caption{Experiment 1 post-study questionnaire results.}
    \label{fig:acc}
\end{figure}

Upon completion of the task, participants rated how much they thought the system had changed across the rounds on a five point Likert scale.
We performed an independent two-way factorial ANOVA on the Likert scale responses, which revealed that those who provided \textit{interactive feedback} thought the system had changed significantly more negatively than those who provided \textit{binary feedback}, with $F(1,103)=3.95$, $p<0.050$, $\eta_G^2 = 0.037$.
\added{No significant effect was observed based on \textit{feedback usage}, nor were any significant interaction effects.}
Results are shown in Figure \ref{fig:changeScaleE}.
% \donald{maybe directly mention something about how it's being carried by the with update condition but wasn't statistically significant}

% \begin{figure*}[tb]
    
%     \centering
%     \begin{subfigure}{.42\textwidth}
%         \captionsetup[subfigure]{justification=centering}
%         \centering
%         \includegraphics[width=\textwidth]{figures/changeScaleECropped.pdf}
%         \caption{Change Caption Goes Here}
%         \label{fig:sub1}
%     \end{subfigure}%
%     \begin{subfigure}{.04\textwidth}
        
%     \end{subfigure}
%     \begin{subfigure}{.42\textwidth}
%         \centering
%         \includegraphics[width=\textwidth]{figures/trustAvgECropped.pdf}
%         \caption{Trust Caption Goes Here}
%         \label{fig:sub2}
%     \end{subfigure}%
%     \begin{subfigure}{.12\textwidth}
%         \centering
%         \includegraphics[width=\textwidth]{figures/E1Legend.pdf}
%         \label{fig:sub4}
%     \end{subfigure}%
    
%     % \caption{Test combined}
%     \label{fig:acc}
% \end{figure*}

% \begin{figure}[tb]
%     \centering
%     \includegraphics[width=0.5\columnwidth]{figures/changeScaleE.pdf}
%     \caption{Extended change scale graph. \TODO{needs real caption}}
%     \label{fig:changeScaleE}
% \end{figure}

\subsubsection{User Trust}
\label{sec:trust_measure}
To measure participants' trust in the system, we asked them to rate their agreement to several statements about different aspects of trust based on an adjusted version of scales proposed by Madsen and Gregor~\cite{madsen2000measuring}.
The following three statements were shown to participants:
\label{subsec:trust1}
\begin{itemize}
    \item The system performs reliably.
    \item The outputs the system produces are as good as that which a highly competent person could produce.
    \item It is easy to follow what the system does.
\end{itemize}

We asked participants to rate their agreement with these trust statements on a seven-point Likert scale.
\added{We analyzed the aggregated responses to the three trust statements with a two-way factorial ANOVA and observed no significant results for either \textit{feedback type} or \textit{feedback usage}, as well as no interaction effects.}
Results are shown in Figure \ref{fig:trustAvgE}.

% \begin{figure}[tb]
%     \centering
%     \includegraphics[width=0.5\columnwidth]{figures/trustAvgE.pdf}
%     \caption{Extended trust agreement scale graph. \TODO{needs real caption}}
%     \label{fig:trustAvgE}
% \end{figure}

\subsection{Summary of Experiment 1 Findings}
Our goal in this experiment was to better understand how \textit{feedback type} and \textit{feedback usage} contributed to the results observed in the initial study~\cite{honeycutt2020soliciting}.
The main finding was that those who gave \textit{interactive feedback} had a more negative perception of the system's change in accuracy over time than those who only gave \textit{binary feedback} \added{(\textbf{\ref{RQ2.1}})}.
No significant results were observed based on the \textit{feedback usage} condition for any of the measures \added{(\textbf{\ref{RQ1}})}.
% These results suggest that the cause for the observed effect was due to the act of providing interactive feedback itself, rather than the belief that the system would update based on that feedback.
\minor{
The primary finding in the original work that this study was based on was that providing interactive HITL feedback lowered user trust and caused underestimation of system accuracy regardless of actual changes in system performance~\cite{honeycutt2020soliciting}.
This result was hypothesized to be caused by an increase in memorability of system errors due to having to spend more time correcting them, resulting in participants  disproportionately remembering those system errors.
Our new findings support this hypothesis, as an effect was observed based on the act of providing feedback and not based on perception of that feedback being used.
}

In the context of this study, the outputs of the system---and consequently the feedback being provided---are of an objective nature; there are a defined number of faces in the image in defined locations.
Because of this, providing feedback in this context is essentially just correcting obvious mistakes from the perspective of the participants.
In prior research that found an increase in trust and usage rates for systems with HITL components over those without, the feedback being provided has involved more subjective system outputs where providing feedback is contextualized as having the ability to make the system behave closer to their own mental model or to their preferences~\cite{rani2022investigating,parra2015user,dietvorst2018overcoming}.
This difference in context could be why the presence of interactive feedback resulted in a negative bias in this study and in \cite{honeycutt2020soliciting}.
\added{
Additionally, we hypothesize that this context is the cause of the overall decrease in perceived accuracy over time regardless of round.
Since the observed system errors were obviously wrong with no nuance or openness of interpretation, seeing such errors resulted in increased distrust over time.
}
In the next section, we present two research studies that explore how a similar study design plays out in a more subjective context.

% \eric{feels like should be more elaboration to help bridge into study 2. a little more about why this is meaningful and interesting from these results alone. and at least a few sentences of mini discussion section to help transition to the next section}

\section{Experiment 2: Text Classification (Fixed Ordering)}
\label{sec:text_study_1}
This section reports a follow-up experiment that mirror the experimental design of the Experiment 1 study but with a text classification context instead of an image context.

\subsection{Research Objectives}
Although we observed a negative bias among those who provided more interactive feedback in both our original object detection study \cite{honeycutt2020soliciting} and the extended Experiment 1 presented in Section \ref{sec:image_study}, other research has found that users of intelligent systems prefer to use systems that allow them to provide more interactive feedback and have increased levels of trust in such systems \cite{dietvorst2018overcoming,rani2022investigating,parra2015user}.
One notable difference between the systems used for these studies that show a positive effect associated with providing feedback and the two experiments that observed a negative bias from providing feedback is the objectivity of the feedback being provided.
In our studies that observed a negative effect, the feedback being provided corresponded to an objectively correct answer; an image region either contains an object or it does not.
Providing interactive user feedback in this context is correcting objective system errors by providing the only correct answer, without nuance or subjectivity.
However, the studies that observed positive effects associated with providing interactive feedback were conducted in contexts where the correctness of an output was more subjective, and the feedback being provided was more about adjusting the behavior of the model to be closer to the user's own mental model.
\added{
By there not being a single objectively correct answer, the act of providing explanation-based feedback is less about having to correct system errors and more akin to making the system's model closer to the user's mental model.
}

With this difference in mind, we looked to create an experiment with the same experimental design as in the first experiment presented in this paper, but in a context with a more subjective task.
We chose text classification---specifically working with text samples from computer science and history textbooks---as it would be familiar to the study populations to be able to provide subjective feedback without requiring any more advanced domain-specific knowledge.

With the goal of building upon the core concepts of the prior work, we aimed to design a new study that addressed similar research questions in a different context.
\added{
Based on the results of \cite{honeycutt2020soliciting} and Experiment 1, we eliminated the independent variable of \textit{change in accuracy over time} and chose to focus instead on solely the presence or absence of different types of feedback.
The relevant research question from Experiment 1 is as follows:
}
\added{
\begin{enumerate}[align=left, label=\textbf{RQ\arabic*:}, ref=RQ\arabic*, start=2]
    \item \label{RQ2.2} Does providing more interactive feedback result in a difference in user trust or perceived system accuracy when compared to only providing accuracy-based feedback?
\end{enumerate}
}

\added{
Additionally, since the integration of HITL approaches into explainable AI systems can be beneficial for improving users' mental models \cite{mosqueira2023human,argall2009survey} as well as providing a mechanism for rich interactive feedback \cite{estivill2022constructing}, we wanted to explore the relationship between effects on user trust when working with explanation-based user feedback.
}
\added{
To that end, we expanded our goals with these additional research questions:
\begin{enumerate}[align=left, label=\textbf{RQ\arabic*:}, ref=RQ\arabic*, start=3]
    \item \label{RQ3} Does user perception of explanation relevance change if the user provides feedback to the system?
    \item \label{RQ4} Does the more subjective context change the effects on user perception?
\end{enumerate}
}

\begin{figure*}[t]
    \centering
    \includegraphics[width=0.99\textwidth]{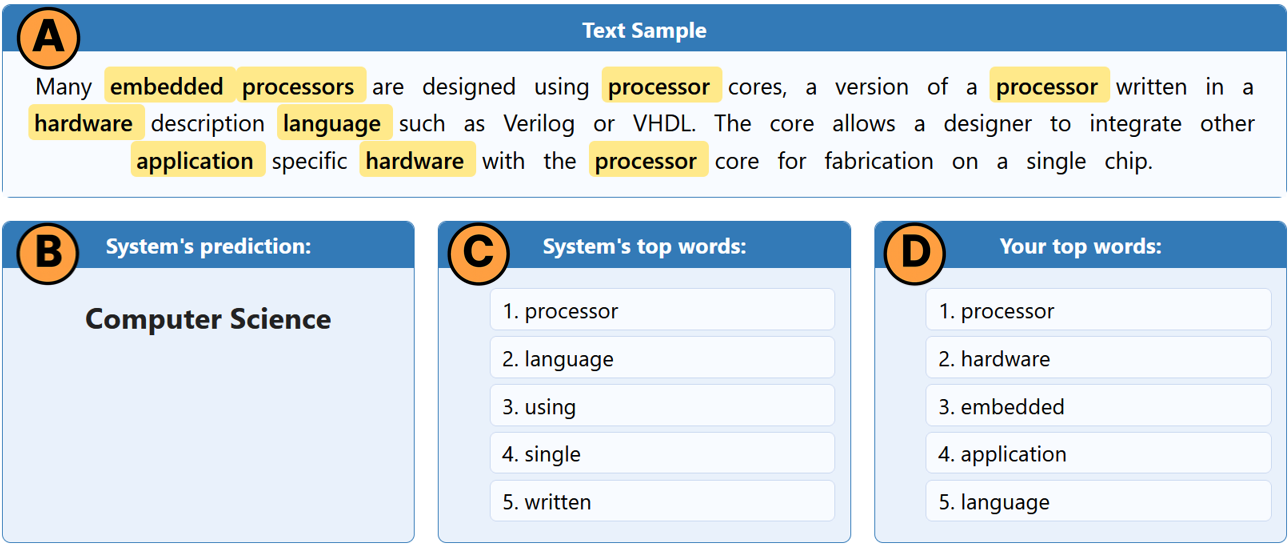}
    \caption{
    \minor{Overview of the Experiment 2 and 3 interface. The system highlights the top words within the text sample (A) that contribute to the classification (B). Participants in the \textit{explanation-based feedback} condition could change the highlights by clicking on the words. The system's original top 5 words are always visible (C) and the participant's updated list can be manually reordered to put the words they perceive as most important at the top (D).}}
    % \footnotemark 
    \label{fig:textInterface}
\end{figure*}

We designed a controlled experiment using a system that classified the subject of text samples with no feedback, decision-based feedback, and explanation-based feedback.
As per our previous results described in this paper, the presence of HITL feedback could reduce user trust and perception of system accuracy by increasing their saliency and causing an availability bias.
Therefore, it could also follow that a similar bias would occur and decrease perceived explanation relevance.
% \donald{unsure of hypothesis here regarding, possible angle:}
% Furthermore, we expect the effect to be more pronounced in users of the system using explanation-based feedback than in those who use the system with decision-based feedback only, as the more involved process of correcting the model's reasoning would make corrections more memorable. 
% \donald{or alternatively:}
However, we hypothesized that using the explanations as a feedback mechanism would offset this effect by reframing feedback in the user's mind from telling the system that it was wrong to explaining their own reasoning back to the system.

\subsection{Method}

This section details of the experimental design and study procedure of Experiment 2.

\subsubsection{Experimental Design}
For the Experiment 2 user study, we developed a text-classification system using a Naive Bayes classifier that predicted the subject of snippets from textbooks\footnote{\url{https://www.kaggle.com/datasets/deepak711/4-subject-data-text-classification/data}}.
The subjects of Computer Science and History were selected as they are easily distinguishable to a general audience.
In addition to the system's prediction, the top five words that most contributed to the class selection were presented as a list, as well as being highlighted in the text.
\added{In order to have a classifier with correctable flaws, we intentionally selected a basic model---Naive Bayes---and  did not randomize the training and testing sets so the classifier was trained on the first 70\% of each textbook and no text from the latter 30\% was included in training.
As a result, the system was not aware of words that were only found in the later chapters of the textbook, which was most noticeable for the History textbook as it was presented in chronological order and the subjects being covered changed significantly based on the time period being discussed.}
Additionally, we intentionally poisoned three words ("hardware", "instruction", and "architecture") by flipping their values for each class.
All three words were strongly weighted towards Computer Science in the initially trained model, but were treated as being related to History by the poisoned model.
Participants completed three rounds of review, with each round containing 15 text samples.
The system prediction accuracy was 80\% across all rounds, which was roughly representative of the overall model performance.

To facilitate our research goals, we used a between-subjects study design with the \textit{level of feedback} as our independent variable with three levels: \textit{no feedback}, \textit{decision-based feedback}, and \textit{explanation-based feedback}.
Participants in the \textit{no feedback} condition were told that they were reviewing the output of a system for quality control and that their responses would be sent to human researchers for later review, with no indication that the system would change based on their input.
\added{
The feedback provided in the \textit{decision-based feedback} condition consisted of only correcting the system's prediction and rating the relevance of the top words on a 5-point Likert scale, whereas the \textit{explanation-based feedback} condition additionally had participants adjusting the highlighted words to create a new list of top relevant words based on their own judgment.
While the classification was objective---each text sample was either about History or Computer Science with no samples that were ambiguous---the feedback about the system's top five words that contributed to the classification was more subjective.
The most relevant words are neither purely objective nor purely subjective; certain words may be more or less related to each subject (e.g., "compiler" is more associated with Computer Science than History) and generic words such as articles and prepositions should not be contributing to the classification one way or another.
However, the specific top five most relevant words in order are not an objective truth like the location of human faces in images as in Experiment 1.
Different people may have different opinions on the most relevant words to explain why a text sample is the topic that it is, whereas the presence and location of a face in an image is not an opinion.
}
In order to keep the system's performance constant for all participants, in the conditions with feedback we chose to only simulate the model updating and incorporating participant feedback rather than using an actual HITL system.
However, participants were informed that the system would update using their feedback between each round of the task.
\added{
As in Experiment \ref{sec:image_study}, this decision to deceive participants to believe that the system was updating based on their feedback when it did not was made to ensure that any observed effects would be due to providing feedback and not because they were interacting with an improved version of the system.
}
\added{
Additionally, Experiment 2 also kept the order of the text samples constant to ensure that each participant saw the same system outputs in the same order, thus providing increased experimental control to prevent possible confounds from different participants experiencing instances in different orders~\cite{nourani2024user,nourani2022importance}.
}
To simulate the updating process for participant believability, a 45 second pause was introduced between each round under the premise of waiting for the model to adjust its classification algorithm based on their responses in the previous round.

\subsubsection{Procedure}
The study was conducted remotely through an online web application with no contact between the researchers and the participants during the task.
Participants were first asked to complete a demographic questionnaire that covered their age, gender identification, educational background, and self-reported experience with ML and artificial intelligence to ensure that there was not a significant difference in the study population between conditions.
This was followed by instructions on how to complete the task and included examples of accurate and inaccurate system predictions.
\added{
All participants were instructed that the goal of the classifier was to predict whether a text sample was from a History or Computer Science textbook, and the classifier provided its top five words in order that contributed to its classification.
For participants who would provide \textit{explanation-based feedback} an additional tutorial was provided to explain how to modify the top words for a given text sample, with a test at the end that required them to demonstrate that they understood how to use the word highlighting system.
}

Following the instructions and tutorial (if applicable), the main task consisted of three rounds of reviewing 15 text samples with corresponding system classifications and the system's top five words for the sample.
\added{
The number of samples in each round was reduced from the 30 images in Experiment 1 to 15 text samples due to the increased amount of time required to process each sample due to the change in domain.
}
After each round, participants were asked to estimate how likely they thought the system was to correctly classify a sample of text from 0-100\%.
For the \textit{decision-based feedback} and \textit{explanation-based feedback} conditions, participants were made to wait 45 seconds and were provided with a reminder that they were waiting for the system to update its classification algorithm based on their previous responses, although no such update was occurring.
After finishing the final round of text samples, participants were asked to complete a questionnaire asking for their level of agreement on a 7-point Likert scale with several statements regarding their level of trust in the system and were given the opportunity to provide any comments they had regarding the study.

\subsubsection{Participants}
We recruited participants through Amazon Mechanical Turk with a requirement for participants to have the Masters qualification, an approval rate of greater than 90\%, and a history of 500 or more successfully completed tasks.
Participants ranged from ages 23--68 and lived in the United States at the time of study completion.
\added{To ensure the quality of participant responses, we measured the percentage of responses for which participants correctly identified whether the classification of a text sample corresponded to a system error or not.
Participants were not included in the results if they had less than $75\%$ accuracy for either correct instances or system errors.}
This study had a total of 149 participants, 5 of whom were removed based on the accuracy criteria.
The remaining 144 participants consisted of 72 males and 72 females. 
Participants took approximately 19 minutes on average to complete the study.

\subsection{Experiment 2 Results}
In this section, we present our study's measures and corresponding statistical test results. 
We report statistical test results along with generalized eta squared ($\eta^2_G$) for effect sizes of ANOVA tests.

\subsubsection{User-Perceived Model Accuracy}
Between each task round, we asked participants to estimate as a percentage how accurate they thought the system's classifications were.
We performed a two-way mixed-design ANOVA with \textit{level of feedback} as a between subjects factor and \textit{task round} (R1, R2, R3) as a within subjects factor. 

\begin{figure}[tb]
    \centering
    % The whole row is forced onto one line
    \makebox[\textwidth][c]{%
        % Left plot
        \begin{subfigure}{0.38\textwidth}
            \centering
            \includegraphics[width=\linewidth]{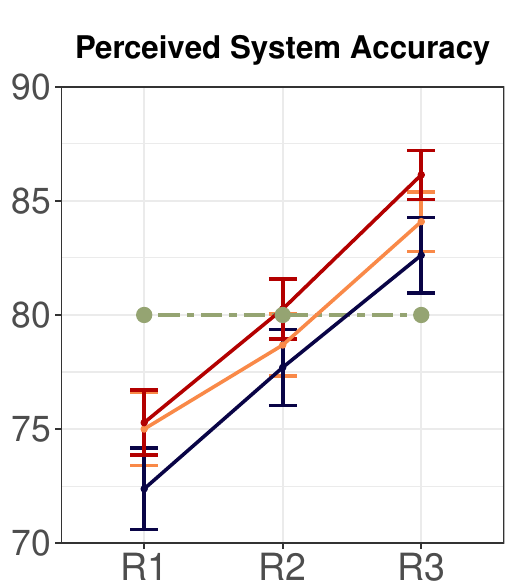}
            \caption{Responses to "How likely do you think the system is to classify a text sample successfully?" (Percentage).}
            \label{fig:linePredTurk}
        \end{subfigure}%
        \hfill
        % Right plot
        \begin{subfigure}{0.38\textwidth}
            \centering
            \includegraphics[width=\linewidth]{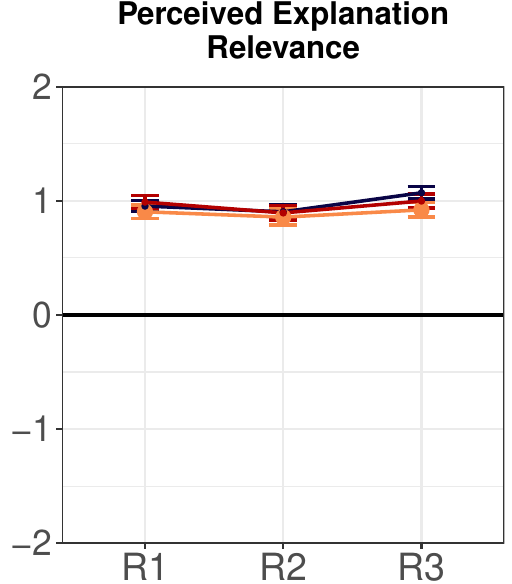}
            \caption{Responses to "How relevant were the system's top words?" on a 5-point Likert scale averaged across each task round.}
            \label{fig:lineRelTurk}
        \end{subfigure}%
        \hfill
        % Legend
        \begin{subfigure}{0.20\textwidth}
            \centering
            \includegraphics[width=\linewidth]{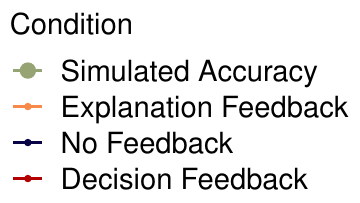}
            \label{fig:legend}
        \end{subfigure}%
    }
    % Optional global caption
    \caption{Experiment 2: Participants' perceived accuracy and ratings of perceived explanation relevance over time, as grouped by the three task rounds (R1, R2, R3) of the study period.}
    \label{fig:acc}
\end{figure}

No significant results were found for the \textit{level of feedback} condition, with $F(2,137)=0.951$, $p=0.389$.
\textit{Task round} was found to have a significant effect, with $F(2,274)=99.328$, $p< 0.001$, $\eta_G^2 = 0.166$.
Post hoc pairwise t-tests with Bonferroni correction revealed significant differences between the task rounds.
Each pair yielded a p-value of $p<0.001$, with the perceived accuracy being higher in each consecutive round (R1 < R2 < R3).
\added{No significant interaction effects were observed between any variables for this measure.}
Results are shown in Figure \ref{fig:linePredTurk}.

% \begin{figure}[h]
%     \centering
%     \includegraphics[width=0.75\columnwidth]{figures/linePredTurk.pdf}
%     \caption{Study 1 accuracy estimation graph. \TODO{needs real caption}}
%     \label{fig:linePredTurk}
% \end{figure}

\subsubsection{User-Perceived Explanation Relevance}
For each text sample, participants rated the system's top words on a five-point Likert scale.
These responses were then averaged for each task round, and we performed a two-way mixed-design ANOVA with \textit{level of feedback} as a between subjects factor and \textit{task round} (R1, R2, R3) as a within subjects factor. 

The \textit{level of feedback} condition was not found to have a statistically significant effect, with $F(2,140)=0.481$, $p=0.619$.
The \textit{task round} had a significant effect observed with $F(2,280)=6.325$, $p=0.002$, $\eta_G^2 = 0.011$.
However, post hoc pairwise t-tests with Bonferroni correction did not reveal significant differences between task rounds.
\added{No significant interaction effects were observed between any variables.}
Results are shown in Figure \ref{fig:lineRelTurk}.

% \begin{figure}[h]
%     \centering
%     \includegraphics[width=0.5\columnwidth]{figures/lineRelTurk.pdf}
%     \caption{First study average perceived relevance graph. \TODO{needs real caption}}
%     \label{fig:lineRelTurk}
% \end{figure}

\subsubsection{User Trust}
\begin{figure}[tb]
    \centering
    % The whole row is forced onto one line
    \makebox[\textwidth][c]{%
        % Left plot
        \begin{subfigure}{0.45\textwidth}
            \centering
            \includegraphics[width=\linewidth]{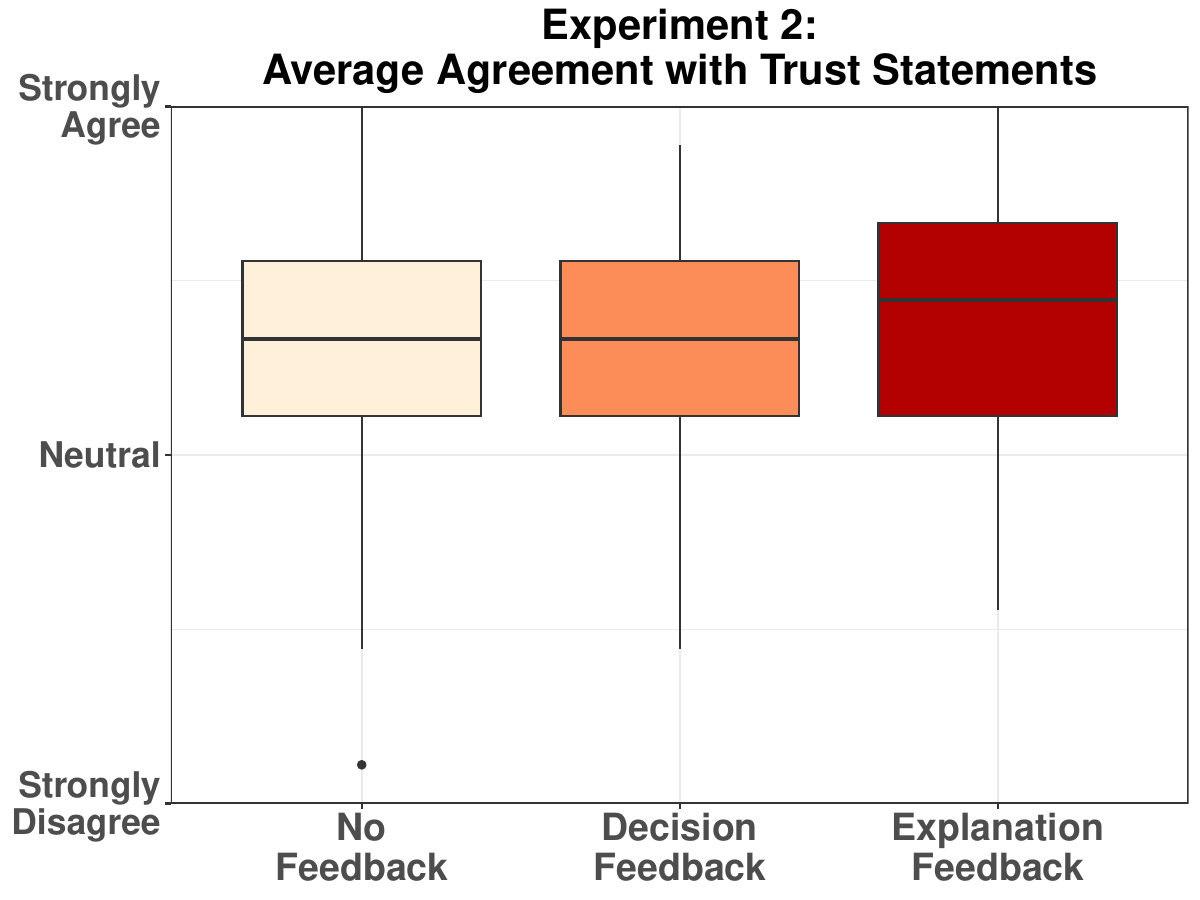}
            \caption{Experiment 2}
            \label{fig:trustAvgTurk}
        \end{subfigure}%
        \hfill
        % Right plot
        \begin{subfigure}{0.45\textwidth}
            \centering
            \includegraphics[width=\linewidth]{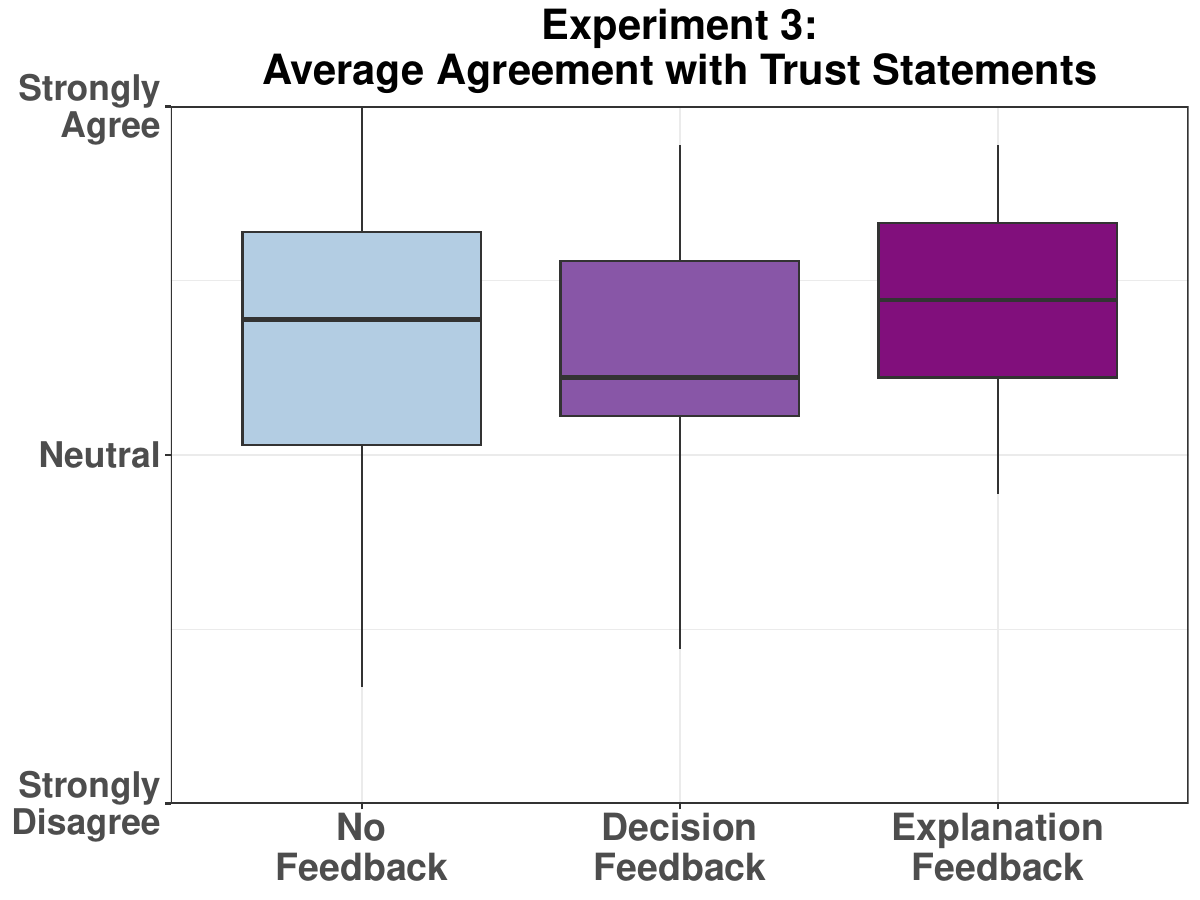}
            \caption{Experiment 3}
            \label{fig:trustAvgSONA}
        \end{subfigure}%
    }
    % Optional global caption
    \caption{Averaged responses to the three trust scale statements described in Section \ref{subsec:trust1} on a 7-point Likert scale.}
    \label{fig:acc}
\end{figure}

At the end of the session, participants rated their agreement with each of a set of trust statements on a seven-point Likert scale.
The same three statements were used as in Experiment 1 (see Section~\ref{sec:trust_measure}), and the aggregate rating was used as a measure for trust.
% The following three statements were shown to all participants, and the aggregate rating was used as a measure for trust:
% \label{subsec:trust2}
% \begin{itemize}
%     \item The system performs reliably.
%     \item The outputs the system produces are as good as that which a highly competent person could produce.
%     \item It is easy to follow what the system does.
% \end{itemize}
For analysis, we performed a one-way ANOVA with \textit{level of feedback} as the between-subjects factor.
No statistically significant result was observed for this measure, with $F(2,141)=0.846$, $p=0.431$.
\added{Additionally, no significant interaction effects were observed between any variables.}
Results are shown in Figure \ref{fig:trustAvgTurk}.

\subsection{Summary of Experiment 2 Findings}
Our goal for Experiment 2 was to explore the effects of providing interactive feedback in a context where the feedback being provided is of a somewhat subjective nature, as contrasted to the more objective detection task of Experiment 1.
Although the Experiment 1 image study described in Section \ref{sec:image_study} and the results from our initial study~\cite{honeycutt2020soliciting} both showed a negative bias toward the perception of system accuracy and a reduced level of trust in the system from providing interactive feedback \added{(\textbf{\ref{RQ2.2}})}, no such effect was observed in the context of text classification in Experiment 2---despite the experimental design approximating the earlier studies. 
\added{
Additionally, no effect was observed on perceived explanation relevance (\textbf{\ref{RQ3}})
}
\minor{
As the experimental design was retained across the object detection and text classification studies, we hypothesize that the level of subjectivity in the context of the feedback mechanism is responsible for the effects observed in \cite{honeycutt2020soliciting} and Experiment 1 in Section \ref{sec:image_study} \added{(\textbf{\ref{RQ4}})}.
}

We hypothesize that when the feedback being provided is of a subjective nature---such as which words are the most relevant to identifying a text snippet's subject---the act of providing feedback is contextualized as modifying the system to behave like one's own mental model.
This is much closer to the contexts of the existing research which has found positive effects on trust and willingness to rely on HITL systems~\cite{rani2022investigating,parra2015user,dietvorst2018overcoming}, leading us to believe that this is why we did not observe the same negative effect as in Experiment 1.
Similarly, we hypothesize that this is also the reason for the observed increase in perceived system accuracy across the rounds of the task.
In the object detection context, providing feedback meant the user was spending most of the time and mental energy associated with providing feedback to correcting objective system errors, resulting in a negative bias towards the system due to those errors being more salient.
In contrast, the more subjective context of the text classification studies had participants actively thinking about their mental model for how to classify the text samples and comparing it to the system's model.
Since the system's accuracy did not change across the rounds of the task, participants perceiving the system to improve in accuracy could indicate that they were adjusting their mental model to be more accepting of the system's reasoning as they spent more time interacting with it.

% Our goal in this extended study was to better understand how \textit{feedback type} and \textit{feedback usage} contributed to the results observed in the initial study.
% The main finding was that those who gave \textit{interactive feedback} had a more negative perception of the system's change in accuracy over time than those who only gave \textit{binary feedback}.
% No significant results were observed based on the \textit{feedback usage} condition for any of the measures.
% These results suggest that the cause for the observed effect in the initial study was due to the act of providing interactive feedback itself, rather than the belief that the system would update based on that feedback.
% Our hypothesis for the primary finding in the original paper---that providing interactive HITL feedback lowered user trust and perception of system accuracy---was that an increase in memorability of system errors caused by having to spend more time correcting system errors could have caused participants to disproportionately remember those system errors, which aligns with our new findings.

\section{Experiment 3: Text Classification (Counterbalanced Ordering)}
\label{sec:text_study_2}
Following the unexpected results of Experiment 2, we sought to deepen the investigation through an additional text classification study following the experimental design of Experiment 2 (Section \ref{sec:text_study_1}).

\subsection{Research Objectives and Experimental Design}

As the results of Experiment 2 showed that participants erroneously believed the system became significantly more accurate across task rounds regardless of condition (Figure \ref{fig:linePredTurk}), we conducted a follow-up study for confirmation. 
Experiment 3 repeats the study design of Experiment 2 but with a counterbalanced ordering of system outputs to ensure that our observations were not caused by the choice of constant ordering of instances in Experiment 2.
\added{
Though the constant ordering in Experiment 2 was an intentional decision based on prior work showing differences and biases due to seeing the same system outputs in different orderings~\cite{nourani2024user,nourani2022importance}, there was a possibility that the specific tested ordering of observed text instances could have (by chance) contributed to the finding of increased perceived accuracy over time. 
To address this potential confound, Experiment 3 counterbalanced the three groups of system outputs so that all possible orders of the text samples and their outputs were represented equally.
}
Other than this change to how participants experienced ordering,  the other aspects of the experimental design (the research objectives, study design, and experimental procedure) remain identical to Experiment 2 as described in Section \ref{sec:text_study_1}.
Participants again provided feedback to the outputs of a text-classification system that predicted the subject of snippets from History and Computer Science textbooks\footnote{\url{https://www.kaggle.com/datasets/deepak711/4-subject-data-text-classification/data}}.
Experiment 3 used the same between-subjects study design with the \textit{level of feedback} as our independent variable with three levels: \textit{no feedback}, \textit{decision-based feedback}, and \textit{explanation-based feedback}, as in Experiment 2.

% \donald{feels a bit weird to just name them without explaining but feels worse to re-explain them or relink the previous section, idk maybe it's fine}
% \eric{even though it's the same, i think this part should still provide bare minimum info about the experimental design...at least name the independent variables and say it used the same text classification task with History and Computer Science}

\subsection{Participants}
For Experiment 3, participants were recruited from the University of Florida Department of Computer and Information Science and Engineering, consisting of both undergraduate and graduate students.
Participants ranged from ages 26--66 and lived in the United States at the time of study completion.
\added{To ensure the quality of participant responses, we measured the percentage of responses for which participants correctly identified whether the classification of a text sample corresponded to a system error or not.
As a quality check, participants were not included in the results if they had less than $75\%$ accuracy for either correct instances or system errors.}
This study had a total of 104 participants, and 10 were removed based on the accuracy criteria.
The remaining 94 participants consisted of 32 females, 60 males, and 2 non-binary.
Participants took approximately 24 minutes on average to complete the study.

\subsection{Experiment 3 Results}

In this section, we present our study's measures and corresponding statistical test results. 
We report statistical test results along with generalized eta squared ($\eta^2_G$) for effect sizes of ANOVA tests.
% \eric{might avoid redundant reporting of same analysis methods, or just leave it. i don't know}
% \donald{I think I prefer it being there as a reminder vs not but am open to dropping it}

\subsubsection{User-Perceived Model Accuracy}

Between each task round, we asked participants to estimate as a percentage how accurate they thought the system's classifications were.
We performed a two-way mixed-design ANOVA with \textit{level of feedback} as a between subjects factor and \textit{task round} (R1, R2, R3) as a within subjects factor. 

\begin{figure}[tb]
    \centering
    % The whole row is forced onto one line
    \makebox[\textwidth][c]{%
        % Left plot
        \begin{subfigure}{0.38\textwidth}
            \centering
            \includegraphics[width=\linewidth]{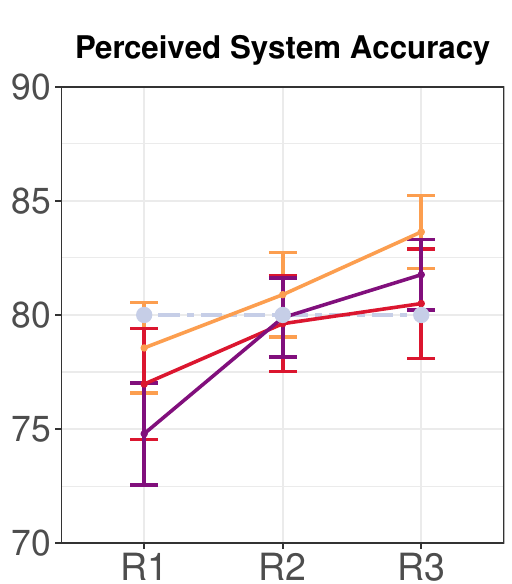}
            \caption{Responses to "How likely do you think the system is to classify a text sample successfully?" (Percentage)}
            \label{fig:linePredSONA}
        \end{subfigure}%
        \hfill
        % Right plot
        \begin{subfigure}{0.38\textwidth}
            \centering
            \includegraphics[width=\linewidth]{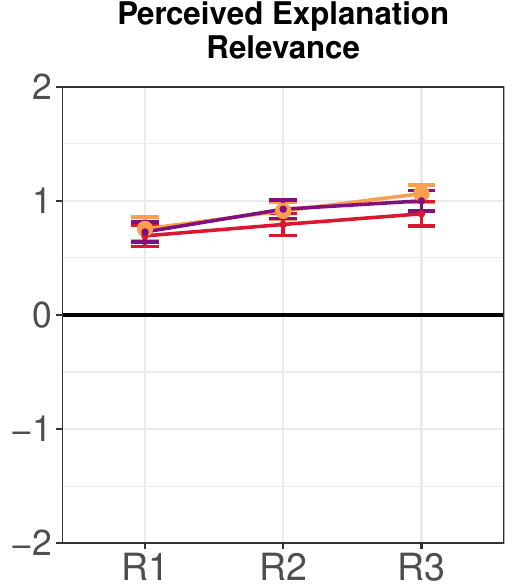}
            \caption{Responses to "How relevant were the system's top words?" on a 5-point Likert scale averaged across each task round.}
            \label{fig:lineRelSONA}
        \end{subfigure}%
        \hfill
        % Legend
        \begin{subfigure}{0.20\textwidth}
            \centering
            \includegraphics[width=\linewidth]{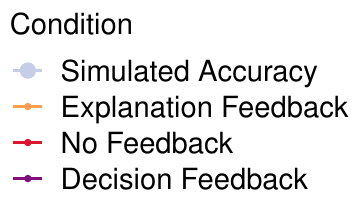}
            \label{fig:legend}
        \end{subfigure}%
    }
    % Optional global caption
    \caption{Experiment 3: Participants' perceived system accuracy and ratings of explanation relevance over time across the three task rounds (R1, R2, R3).}
    \label{fig:acc}
\end{figure}

The \textit{level of feedback} condition did not have significant results, with $F(2,91)=0.871$, $p=0.422$.
A significant effect was observed based on \textit{task round}, with $F(2,182)=14.194$, $p<0.001$, $\eta_G^2 = 0.045$.
Post hoc pairwise t-tests with Bonferroni correction revealed significant differences between the task rounds.
Specifically, the comparison between R1 and R3 yielded a p-value of $p < 0.0015$, with user perception of system accuracy in R3 being higher than in R1.
The other paired comparisons did not yield statistically significant results.
\added{Additionally, no significant interaction effects were observed between any variables.}
Results are shown in Figure \ref{fig:linePredSONA}.

\subsubsection{User-Perceived Explanation Relevance}
For each text sample, participants rated the system's top words on a five-point Likert scale.
These responses were then averaged for each task round, and we performed a two-way mixed-design ANOVA with \textit{level of feedback} as a between subjects factor and \textit{task round} (R1, R2, R3) as a within subjects factor. 

\textit{Level of feedback} was not found to have a statistically significant effect, with $F(2,87)=0.853$, $p=0.430$.
A significant effect was observed based on \textit{task round}, with $F(2,174)=13.079$, $p< 0.001$, $\eta_G^2 = 0.057$
Post hoc pairwise t-tests with Bonferroni correction revealed significant differences between the task rounds.
Specifically, the comparison between R1 and R3 yielded a p-value of $p < 0.001$, with user perception of explanation relevance in R3 being higher than in R1.
The other paired comparisons did not yield statistically significant results.
\added{No significant interaction effects were observed between any variables.}
Results are shown in Figure \ref{fig:lineRelSONA}.

\subsubsection{User Trust}
At the end of the session, we asked participants to rate their agreement with each of a set of trust statements on a seven-point Likert scale.
The same three statements were used as in Experiments 1 and 2 (described in Section~\ref{sec:trust_measure}), and the aggregate rating was used as a measure for trust.
% The following three statements were shown to all participants, and the aggregate rating was used as a measure for trust:
% \begin{itemize}
%     \item The system performs reliably.
%     \item The outputs the system produces are as good as that which a highly competent person could produce.
%     \item It is easy to follow what the system does.
% \end{itemize}
We performed a one-way ANOVA with \textit{level of feedback} as the between-subjects factor.
No statistically significant result was observed for this measure, with $F(2,101)=1.509$, $p=0.226$.
\added{Additionally, no significant interaction effects were observed between any variables for this measure.}
Results are shown in Figure \ref{fig:trustAvgSONA}.

\subsection{Summary of Experiment 3 Findings}

The main goal of Experiment 3 was to ensure that the perceived increase in accuracy over time observed in Experiment 2 was not due to something specific with the fixed ordering of text samples.
Additionally, this study used Computer Science students as participants whereas the study population from Experiment 2 was crowdsource workers on Amazon Mechanical Turk.

There was no significant effect based on \textit{level of feedback} for any metrics.
However, participants in all conditions perceived the system to become more accurate over time and have more relevant explanations by the final round of the task than they did for the first round. 
The only difference from the Experiment 2 results is the significant result for perceived explanation relevance.
This result is similar to the result for perceived accuracy in that participants perceived the system to be improving over time when it was not.
\added{
These results support that the increase in perceived accuracy across the task rounds in Experiment 2 was not due to the fixed ordering.
Additionally, the lack of an effect based on \textit{level of feedback} contributes more evidence that there is a difference due to task context between the object detection and text classification studies.
The only major difference between the results of Experiment 2 and Experiment 3 was the average task length---19 minutes on average for Experiment 2 and 24 minutes on average for Experiment 3.
This is likely due to the differences between the study populations.
The crowdsource workers from Amazon Mechanical Turk used in Experiment 2 were paid a flat rate based on the expected length of the task, so the more tasks they can complete in a given period of time the higher their effective hourly pay rate becomes.
While the students in Experiment 3 were also given a flat amount of extra credit, the total amount of extra credit they could receive was fixed regardless of how quickly they completed the task.
}

In the next section, we will discuss the results of all three experiments together along with a selection of the most relevant related work.

\section{Discussion}
\label{sec:discussion}
In this section, we discuss the results of the three experiments in the context of the existing body of research on the effects of providing users with HITL mechanisms.
We also consider limitations of our work and opportunities for continued research.

\subsection{Effects of Interactive User Control on User Perception}
% \eric{needs a little bit of summary of what was learned from all three studies together. what was the overall finding? what is important about our studies before you start talking about others?}
% \eric{this section just looks like more related work. it's missing connection to your studies in the paper}
Several studies have observed positive effects on user behavior and system perception by providing the opportunity to give HITL feedback.
\added{
Multiple studies have shown participants to have a stated preference towards versions of intelligent systems that allow them to adjust model behavior to be closer to how they would complete the system's task themselves~\cite{parra2015user,rani2022investigating,jin2017different,stumpf2008integrating,he2016interactive}.
}
This stated preference is reflected in users' behavior; while users will stop using intelligent systems they perceive to be flawed---even when their use would still improve performance \cite{dietvorst2015algorithm}---this aversion can be alleviated through the presence of even a minimal level of user control \cite{dietvorst2018overcoming}.
When using an intelligent citation recommendation system, users who were allowed to adjust the weights of the recommender perceived the same recommendations as more relevant and displayed a pattern of accepting recommendations immediately after making a HITL adjustment~\cite{parra2015user}.

However, the presence of these HITL feedback mechanisms is not without cost.
Several studies have shown an increase in cognitive load resulting from an increased level of user control, which can result in decreased levels of engagement with the HITL mechanisms despite their effectiveness \cite{rani2022investigating,jin2017different,ehrmann2022evaluating}.
When the HITL feedback mechanism is purely corrective---treating the user as an active learning oracle as we did in Experiment 1---it can result in users underrating the accuracy of the system and trusting the system less \cite{honeycutt2020soliciting,iuzzolino2020automation}.
Even when the HITL mechanisms are viewed as a positive addition by the user, those feelings can be reversed if they feel that their feedback is not being reflected in meaningful behavioral change throughout the system, resulting in a higher level of frustration with the system than if the system had no HITL options at all \cite{stumpf2008integrating}.

\subsection{Importance of Feedback Objectivity}
\added{
Both the results of the experiments presented in this paper and the existing body of literature on the effects of providing HITL feedback on user perception suggest that the context of the system that feedback is being provided to contributes to user perception of system accuracy.
}
In the context of object detection in images from Experiment 1, participants were providing feedback of an objective nature; a region of a picture either contains a human face or it does not.
There was no room for individual preference, but rather the system is either demonstrating a correct output or an incorrect output and the participants were tasked with correcting the errors.
The main finding of this experiment was that participants who gave \textit{interactive feedback} perceived the system to have a greater decrease in accuracy across the rounds of the task than those who only gave \textit{binary feedback} (Figure \ref{fig:changeScaleE})
Additionally, we observed that participants across all conditions perceived a significant decrease in system accuracy throughout the task, even though the system accuracy remained constant (Figure \ref{fig:ELine}).
In contrast, the text classification context used in Experiments 2 and 3 provide a more subjective form of feedback.
Although the true classification of a text sample's topic is objectively known, identifying which words most contribute to that classification is somewhat subjective.
These studies did not observe a negative bias based on feedback interactivity, and participants in all conditions believed that the accuracy of the system improved across the duration of the task (Figure \ref{fig:linePredSONA} and Figure \ref{fig:linePredTurk}).

\added{
When the feedback being provided is of an objective nature---treating the user as an active learning oracle as in the image context of Experiment 1 and active learning literature~\cite{iuzzolino2020automation}---the act of providing feedback is contextualized as having to correct a faulty system and results in perceiving the system to be worse than it actually is due to the increased focus on system errors.
In contrast, when the feedback is of a subjective nature---as in the text context of Experiments 2 and 3 where the most relevant words are open to interpretation---it can be contextualized as having the ability to make the system behave closer to the user's own mental model, which is closer to the reasons for preferring systems with HITL components observed by prior research~\cite{parra2015user}.
In studies where users have been found to report increased satisfaction or exhibited preferences towards systems with HITL feedback, the feedback mechanisms have predominantly been collaborative systems framed as providing user control based on their preferences \cite{rani2022investigating,parra2015user,dietvorst2018overcoming}.
The feedback in these systems is even more subjective than the annotative feedback in Experiments 2 and 3, and stronger effects are observed in those contexts.
We hypothesize that this distinction is responsible for the differences observed across the different research in this field.
}

Furthermore, we hypothesize that the differences in the explanations provided across the different study contexts contributed to participants in the image classification study perceiving the system to get worse over time, while participants in the text classification studies found the system to become more accurate over time, despite the accuracy of the systems remaining constant in all three studies.
In the image classification context, the objectivity of the presence of faces within the bounding boxes contributed to system errors seeming obvious.
The location of faces in an image is relatively obvious to identify as a human, and bounding boxes alone do not provide much insight into the reason behind model errors.
In the text classification studies, we provided participants with the top five words that contributed to the classification of each text sample.
Since there is no objectively correct set of five words which most contribute to a text sample's topic, there is room for both the user to provide feedback that feels like they are making the system behave closer to their mental model and for the user to adjust their mental model to be closer to the system's.
This could be the cause of participants finding the system to be more accurate over time in the more subjective context of our text classification studies.

\subsection{Design Implications}
A potential reason to include HITL components is to crowd-source model improvements, giving end users the ability to correct objective system errors.
Although this may be a beneficial feature for both model improvement and user control, this research suggests that treating the end user as an active learning oracle in this manner could actually result in a negative bias towards their perception of the system's accuracy over time.
Overly distrusting an intelligent system in this manner can result in self-reliance for critical decisions, resulting in higher rates of failure than if they trusted the system at an appropriate level~\cite{parasuraman1997humans}.
This could be offset by including types of feedback that users feel allow them to influence model behavior rather than just having them correct objective mistakes.
Additionally, providing robust explanations of model behavior can help calibrate an appropriate level of trust~\cite{teso2018should,ribeiro2016should}.
System explanations can also act as a natural mechanism for more detailed user control, helping the act of providing feedback to be more about changing model behavior rather than simply labeling mistakes.
However, over-complicating the feedback mechanisms can result in an increase in cognitive load that may cause less experienced users to engage with the system less~\cite{rani2022investigating,jin2017different,ehrmann2022evaluating}.
Finding the appropriate level of explanations and control to provide to different types of users is important to ensure an appropriate level of engagement and reliance.

Designers of intelligent systems should also consider whether they are in a context where users will be able to objectively identify system errors or whether errors are more subjective and difficult to identify.
In the image classification study where system outputs were fully objective, we observed a negative bias towards the system over time; in the text classification studies where system outputs were more subjective we saw the opposite effect.
The ideal behavior would be for a user of an intelligent system to base their decision to rely on a system off of examining the system's outputs and making an evaluation based on its performance over time~\cite{hoffman2013trust}.
However, users do not always engage in this behavior, particularly if they are not experts in the domain of the system~\cite{nourani2020Role}.
The findings of this paper suggest that when system errors feel objectively wrong, it results in overly distrusting the system.
On the other hand, there is also a risk of over-reliance---also known as automation bias~\cite{goddard2012automation}---if the user has difficulty identifying system errors~\cite{lee2004trust,hoffman2013trust,nourani2020investigating}.
Ensuring that even less experienced users can make informed decisions about the trustworthiness of AI systems is critical to avoid improper levels of reliance.

\subsection{Limitations and Future Work}
\minor{
The main limitation of this work is that neither system studied in this paper actually utilized the participants' feedback to update the system, instead relying on deceiving the participants into thinking it was.
}
Although this provided a necessary level of experimental control, it increases the gap between this research and the real world scenarios our findings could be applied to.
Other studies have utilized systems containing actual HITL updates, but they often fail to distinguish between the effects of providing feedback and the effects of interacting with the resulting improved system \cite{parra2015user,jin2017different}.
There is a clear need for further exploration into studying the effects of providing HITL feedback on system perception and user behaviors in increasingly realistic contexts.
Additionally, both studies presented in this paper used mandatory feedback, while most related work focuses on optional feedback---a more common real-world scenario.
A potential avenue for further research would be to explicitly explore if there is a difference in the observed effects when working with mandatory versus optional feedback.

This research contributed empirical evidence that asking users of intelligent systems to provide corrective feedback in the style of an active learning oracle can result in underrating system accuracy and distrust towards the system.
We also showed that this effect was not observed when repeating the experiment in the context of asking the user to provide more subjective feedback.
\added{
However, these were separate studies and subjectivity was not a variable that was directly controlled for.
Additionally, the more subjective text case has more of an element of explainability than the more objective image case---a bounding box around a face is less of an explanation than it is part of the classification---which may have also contributed to the differences in observed effects.
While our experiments and prior work support the interpretation that level of subjectivity is contributing to the differences in observed results across studies, future research that directly explores the effects of different levels of subjectivity on the user perception is needed.
}

\added{
The studies presented in this paper all observed significant changes in perception of accuracy over time, which was not what the studies were directly designed to explore.
Future research about the change in perception of intelligent system accuracy over time could consider more task rounds or longer durations, potentially focusing on extended usage over multiple sessions.
They could also utilize more involved tasks or observe real world intelligent system use where the participants are more invested in the system outcomes.
}

\section{Conclusion}
\label{sec:conclusion}

Including HITL feedback in an intelligent system can be beneficial, both in terms of improved model performance and providing users with agency over system behavior.
When it comes to its effects on user satisfaction, perception of system performance, and user behaviors, results can wildly vary.
We presented three experiments that explored how different feedback contexts can change the effects that providing such feedback has on users of HITL intelligent systems.
From Experiment 1, we found that in cases where the accuracy of a given system output feels objective, users distrust the system through continued use.
However, in a context where the outputs of a system are judged more subjectively (Experiments 2 and 3), participants perceived that the system improved over time despite no change in its true accuracy.
This suggests the perceived objectivity of system accuracy could contribute to both distrust and mistrust.
Furthermore, the results showed that when participants are treated as an active learning oracle and asked to provide objective feedback about the accuracy of the system and correct its errors, they distrust the system and underrate its accuracy.
However, no distrust was observed when participants were given a similar task but asked to provide feedback that was subjective in nature.
These results suggest that developers who wish to incorporate HITL mechanisms without into their systems without negatively biasing their users should focus on using it to enhance user agency rather than as a method of objective data collection.

\section{Acknowledgements}
Anonymized for review
% This work was supported by the DARPA Explainable Artificial Intelligence (XAI) Program under contract number N66001-17-2-4032 and by NSF award 1900767.

\bibliographystyle{acm}
\bibliography{bibliography.bib}

\end{document}